\documentclass[12pt]{article}
\usepackage{color}
\usepackage{graphicx}
\usepackage{times}
\usepackage{amssymb, amsmath,color}
 \usepackage[square,sort,comma,numbers]{natbib}
\setcitestyle{square}
\usepackage{lineno}

\title{Recent north magnetic pole acceleration towards Siberia caused by flux lobe elongation}

\author
{Philip W. Livermore,$^{1\ast}$, Christopher C. Finlay $^{2}$, Matthew Bayliff $^{1}$\\
\\
\normalsize{$^{1}$School of Earth and Environment, University of Leeds, Leeds, LS2 9JT, UK,}\\
\normalsize{$^{2}$DTU Space, Technical University of Denmark, 2800 Kgs. Lyngby, Copenhagen, Denmark}\\
\\
\normalsize{$^\ast$To whom correspondence should be addressed; E-mail:  p.w.livermore@leeds.ac.uk.}
}
\date{}

\begin{document} 


\maketitle 

\begin{abstract}
The wandering of Earth's north magnetic pole, the location where the magnetic field points vertically downwards, has long been a topic of scientific fascination.
Since the first in-situ measurements in 1831 of its location in the Canadian arctic, the pole has drifted inexorably towards Siberia, accelerating between 1990 and 2005 from its historic speed of 0-15 km/yr to its present speed of 50-60 km/yr. In late October 2017 the north magnetic pole crossed the international date line, passing within 390 km of the geographic pole, and is now moving southwards. 
Here we show that over the last two decades the position of the north magnetic pole has been largely determined by two large-scale lobes of negative magnetic flux on the core-mantle-boundary under Canada and Siberia. Localised modelling shows that elongation of the Canadian lobe, likely caused by an alteration in the pattern of core-flow between 1970 and 1999, significantly weakened its signature on Earth's surface causing the pole to accelerate towards Siberia.  A range of simple models that capture this process indicate that over the next decade the north magnetic pole will continue on its current trajectory travelling a further 390-660 km towards Siberia.

\end{abstract}
Historical determinations of the pole position, for example by Ross in 1831 \cite{Ross_1834}, and later by Amundsen in 1904 \cite{Amundsen_1908}, relied on ground surveys, searching for the location where the horizontal component of magnetic field $H$ was zero and a magnetic needle pointed directly down to the center of the Earth \cite{Good_1991}.  Such direct determinations are difficult, especially if the pole position is not on land and because of field fluctuations due to currents in the high latitude ionosphere \cite{Newitt_etal_2009}.  More recently the magnetic pole position has been determined from global models of the geomagnetic field \cite{Thebault_etal_2015a} built using  measurements made by both satellites and by a network of ground observatories. The accuracy of such pole determinations, which depends on the quality and distribution of the contributing observations along with the ability to remove the external magnetic field, has steadily improved over time; since 1999 there has been continuous monitoring of the geomagnetic field from space by a series of dedicated satellite missions, most recently the Swarm mission \cite{Friis-Christensen_etal_2006}. In Fig 1a. we show the path of the pole since 1840 from the COV-OBS.x1 \cite{Gillet_etal_2015a} and CHAOS-6-x8 \cite{Finlay_etal_2016} geomagnetic field models alongside in-situ historical measurements. The location of the magnetic pole is a characteristic of the core-generated magnetic field that is spherically-radially attenuated through the mantle, which may be considered as an electrical insulator on the time-scales of relevance here.  The magnetic pole's position is thus only an indirect indicator of the state of Earth's dynamo. However the specific geometry of the magnetic field on Earth's surface is of broad societal importance, as was demonstrated recently by the need for a high-profile irregular update in 2019 of the world magnetic model used for navigation in many mobile devices \citep{Chulliat_etal_2019}.

\subsection*{Recent movement of the north magnetic pole}
Compared with its meandering position prior to the 1970s, over the past 50 years the north magnetic pole has travelled along a remarkably linear path that is unprecedented in the recent historical record \cite{Hope_1957,Olsen_Mandea_2007,Korte_Mandea_2008}, guided along a trough of low horizontal field \cite{Hope_1957, Mandea_Dormy_2003}. Using high-resolution geomagnetic data from the past two decades \cite{Finlay_etal_2016}, Figs 2a,d show that this trough connects two patches of strong radial magnetic field at high latitude centred on Canada and Siberia. The importance of these two patches in determining the structure of the field close to the north magnetic pole has been well known for several centuries \cite{Hansteen_1819}. 
Both the path of the north magnetic pole and the crucial Canadian and Siberian magnetic patches are characteristics of the large-scale field \cite{Korte_Mandea_2008}, already evident when the field is truncated at spherical harmonic degree $l$=6 (Figs 2b,e). Considered in isolation from the remainder of the global field, each Earth-surface patch of strong radial field would define a magnetic dip pole close to its centre point. The present two-patch structure of the high latitude geomagnetic field then defines two ends of a linear conduit of near vertical field along which the north magnetic pole can readily travel. 

Between 1999 and 2019, the Siberian patch showed a slight intensification from a minimum value of -60.5 to -60.6 $\mu$T, while the Canadian patch decreased significantly in absolute value from a minimum of -59.6 to -58.0 $\mu$T (Fig 2a,d). Together, these caused the direction of travel of the north magnetic pole to be towards Siberia. 

Although the magnetic field on Earth's surface is linearly related to the structure of the field on the core-mantle boundary (CMB), the geometric attenuation through the mantle means that this relationship is not a simple mapping. For example, the north magnetic pole does not correspond to a location on the CMB where the horizontal field vanishes, but rather reflects a non-local averaging of the field as shown in Figures 2b,c and 2e,f.
The important Canadian and Siberian surface patches are also spatial averages over regions dominated by the large-scale lobes of intense magnetic flux underneath Canada and Siberia on the CMB that are themselves fundamental features of the geodynamo process (Figs 2c,f) \cite{Bloxham_Gubbins_85}.
We find that the time-dependent position of the pole along the conduit is largely governed by a balance or tug-of-war between the competing influences of the Canadian and Siberian lobes on the CMB.  The angular offset between the pole and these controlling flux lobes at mid to high latitudes (50 - 70$^\circ$N) is in accord with the relevant Green's functions for Laplace's equation under Neumann boundary conditions \citep{Chulliat_etal_2010,Gubbins_Roberts_83}. 

\subsection*{Localised flux lobe elongation}
We now probe the physical mechanism that underpins the recent shift in balance between the two flux lobes.
Changes in the CMB radial magnetic field over 1999-2019 (movie S1) show that the Canadian flux lobe (marked A, Fig 3c) elongated longitudinally and divided into two smaller joined lobes (A' and B) within the marked wedge (Fig 3a). Although lobe B has a higher intensity compared to lobe A, importantly the spatial lengthscale of the magnetic field within the wedge has decreased. The transfer of magnetic field from large to smaller scales caused the weakening of the Canadian patch at Earth's surface because smaller scales attenuate faster through the mantle with distance from the source. At the same time the increasing proximity of lobe B to the Siberian lobe enhanced the Siberian surface patch (Fig 3d).
To demonstrate that this elongation effect is the primary cause of the recent north magnetic pole movement, we performed a numerical experiment where we isolated geomagnetic variation over the period 1999-2019 to within the wedge (Figs 3a,c), the geomagnetic field being held fixed at its 1999 structure elsewhere, and calculated the geomagnetic signature on the Earth's surface (see methods). 
This simple model reproduces the weakening of the large-scale part of the Canadian flux lobe at the CMB (Fig 3b) and the concomitant weakening of the Canadian patch at Earth's surface (Fig 3d), in accord with Fig 2; it also reproduces the growth of the Siberian surface patch. Furthermore, it accounts for 961 km of the 1104 km (87\%) distance travelled by the north magnetic pole over 1999-2019.
In a similar vein, we conducted additional numerical experiments (see figs S1, S2 and methods) to test two other localised mechanisms previously proposed to explain the recent north magnetic pole movement: those of intense geomagnetic secular variation under the New Siberian Islands \cite{Chulliat_etal_2010} and the influence of a polar reversed-flux-patch on the CMB \cite{Olsen_Mandea_2007}. Both of these hypotheses produce only small movements of the pole (travelling respectively 142 km and 16 km over 1999-2019). Prior to 1990, and at least as far back as 1940 (movie S2), the COV-OBS.x1 geomagnetic model shows that the Canadian flux lobe was quasi-stable, consistent with the slowly moving magnetic pole. In the 1990s, vigorous elongation leading up to the flux lobe splitting post 1999 resulted in the observed rapid change in speed of the north magnetic pole.

\subsection*{Interpretation in terms of core-flow}
Time variation of the geomagnetic field arises through a combination of core-flow and magnetic diffusion. The reconfiguration of the Canadian flux lobe requires a change in the signature in either or both of these two effects within the core under Canada, although inference of any single underlying  dynamical process is non-unique. Here we base our interpretation on the frozen-flux assumption which asserts that over decadal timescales the impact of core-flow is likely dominant \cite{Roberts_Scott_65}, and is consistent with the formation and advection of lobe B (Fig 3a).
Fig 4a-c shows snapshots of the radial magnetic field with streamlines showing direction and magnitude of the large-scale core surface flow in 1970, 1999 and 2017, depicting flow changes in this region during the acceleration phase of the north magnetic pole. The presented flow models are the ensemble means of a series of flows inferred by probabilistic inversions of both ground-based observatory and satellite data, with a parameterisation of the unknown magnetic diffusion and sub-grid scale induction processes \cite{Barrois_etal_2017, Barrois_etal_2018, Barrois_etal_2019}. 
In 1970, an intense large-scale flow transported magnetic flux northwards under the east-coast of North America, connecting to a polar westwards flow around a section of the inner-core tangent cylinder. Importantly, only a small part of the northward flow at that time passed through the Canadian flux lobe. By 1999 the flow had altered into a broad trans-North-America stream that converged and strengthened under Alaska: this differential velocity was efficient at elongating (by stretching) the Canadian lobe westwards.  By 2017 the flow under Alaska had further strengthened, advection and further stretching acting to separate the Canadian lobe into two pieces. Our interpretation based on the presented ensemble mean flow is reinforced by the fact that the basic sequence of events described above occurs in all flow ensemble members.
 
The strengthening azimuthal flow under the Bering Straits, a key part of the core-flow changes described above, may also be associated with the appearance of an intense tangent-cylinder jet in this region, which has a clear observational signature in the small-scale magnetic field (above spherical harmonic degree 11) after 2004 \cite{Livermore_etal_2017}. However, such a tangent cylinder jet is in itself too localized at high-latitude to be responsible for the elongation of the Canadian lobe in the 1990s that caused the rapid acceleration of the north magnetic pole. Instead it seems that alteration in the global gyre structure \cite{Pais_Jault_2008,Gillet_etal_2015b} beneath North America began the elongation and contemporaneous north magnetic pole acceleration.

\subsection*{Future predictions and historical perspectives}
Fig 1a shows a prediction of the future north magnetic pole position from a variety of models: linear extrapolations from 2019 of the World Magnetic Model (v2) \cite{Chulliat_etal_2019} and CHAOS-6-x8 \cite{Finlay_etal_2016}, and predictions based on the two end-member processes generating geomagnetic secular variation, frozen-flux induction and pure magnetic diffusion (see methods). All the models are based on recently observed secular variation including the elongation of the Canadian flux lobe, and all predict a continuation of the current trajectory of the pole, with the greatest change in position being from one flow ensemble member (660 km) and the minimum change in position from the World Magnetic Model (v2) (390 km).

Will the north magnetic pole ever return to Canada? Given the delicate balance between the Canadian and Siberian flux lobes controlling the position of the pole along the trough of weak horizontal field, it would take only a minor readjustment of the present configuration to reverse the current trend. Predictions of the magnetic field over decade to century timescales are on the horizon using data assimilation methods \cite{Aubert_2015,Tangborn_Kuang_2018,Sanchez_etal_2019}, but these are still under development and for now it is most informative to look at its past behaviour as a guide. Reconstructions of the historical and archeomagnetic field over the past few thousand years are inherently smoothed in time and based on sparse data, but nevertheless can resolve the large-scale field patches that control the location of the magnetic north pole. These reconstructions show that although the northern hemisphere has largely been dominated by two flux patches, occasionally a three-patch structure has arisen which would have had an effect on the pole's position \citep{Nilsson_etal_2014,Panovska_etal_2018}. Over the last 400 years, the pole has meandered quasi-stably around northern Canada, but over the last 7000 years it seems to have chaotically moved around the geographic pole, showing no preferred location \cite{Korte_Mandea_2008}. Analogues of the recent acceleration may have occurred at 4500 BC and 1300 BC when the speed reached about 3-4 times the average seen in these reconstructions. The most recent of these events coincided with the pole moving towards Siberia (from a region close to Svalbard) where it remained stable for several hundred years. 
For now, a conclusive answer to the future location of the north magnetic pole will have to await detailed monitoring of the geomagnetic field from the Earth's surface and space in the coming years.

\clearpage

\section*{Methods}
The isolation of geomagnetic secular variation in specific regions on the CMB as shown in Figs 3, S1 and S2 is achieved using a physical grid: inside the shown wedge the radial component of the geomagnetic field is allowed to evolve, whereas outside it is frozen at its initial state. We transform to an equivalent divergence-free magnetic-potential representation based on spherical harmonics, which allows upward continuation of the magnetic field to the Earth's surface. The latitude-longitude grid has $L+1$ Gauss-Legendre points in colatitude, and $3L+1$ equally spaced points in longitude, where the maximum spherical harmonic degree is $L=13$.  Note that any monopolar component or discontinuities caused by adjoining two distinct magnetic field structures are removed by the projection adopted.

To predict the north magnetic pole position using the large-scale flow ensemble of \cite{Barrois_etal_2018,Barrois_etal_2019}, for each ensemble member all spherical harmonic flow coefficients are extrapolated 2019-2029 using a simple linear best fit through their values from 2014-2018. The rate of change of geomagnetic field is then computed from the induction equation using the time-dependent large-scale flow along with a static correction term. The geomagnetic field is then evolved through time using a first order time-stepping scheme and the position of the north magnetic pole evaluated using a descent method in the horizontal magnitude. The correction term is chosen so that the Gauss coefficients (to degree 13) of the modelled rate of change of geomagnetic field at 2019 match those from CHAOS-6-x8.  Its static nature relies upon on the assumption that both diffusion, and any small-small scale interactions not captured in the large scale flow models, are time-independent over a 10-year period. A purely-diffusive prediction is based on the model of \cite{Metman_etal_2019}, in which a magnetic field diffuses from its initial state. The model is described by two radial basis functions for each poloidal spherical harmonic mode up to a maximum spherical harmonic degree 13. The coefficients describing the initial field (here taken to be in 2014) are chosen by fitting to CHAOS-6-x8 over the time period 2014-19. The model is then evolved beyond 2019 according to the diffusion equation; over this time period it differs from the linear extrapolation of CHAOS-6-x8. Note that this procedure is not sensitive to the specific choice of time window: a model fit over 2018-19 from an initial state in 2018 (not shown) is visually almost indistinguishable from that fit over 2014-19.
  

\begin{thebibliography}{10}
\expandafter\ifx\csname url\endcsname\relax
  \def\url#1{\texttt{#1}}\fi
\expandafter\ifx\csname urlprefix\endcsname\relax\def\urlprefix{URL }\fi
\providecommand{\bibinfo}[2]{#2}
\providecommand{\eprint}[2][]{\url{#2}}

\bibitem{Ross_1834}
\bibinfo{author}{Ross, J.~C.}
\newblock \bibinfo{title}{{On the Position of the North Magnetic Pole}}.
\newblock \emph{\bibinfo{journal}{Phil. Trans. R. Soc. A}}
  \textbf{\bibinfo{volume}{124}}, \bibinfo{pages}{47--52}
  (\bibinfo{year}{1834}).

\bibitem{Amundsen_1908}
\bibinfo{author}{Amundsen, R.}
\newblock \emph{\bibinfo{title}{{The Northwest Passage}}}
  (\bibinfo{publisher}{{Archibald Constable \& Co. Ltd}},
  \bibinfo{address}{London}, \bibinfo{year}{1908}).

\bibitem{Good_1991}
\bibinfo{author}{Good, G.}
\newblock \bibinfo{title}{{Follow the needle: seeking the magnetic poles}}.
\newblock \emph{\bibinfo{journal}{Earth Sciences History}}
  \textbf{\bibinfo{volume}{10}}, \bibinfo{pages}{154--167}
  (\bibinfo{year}{1991}).

\bibitem{Newitt_etal_2009}
\bibinfo{author}{Newitt, L.~R.}, \bibinfo{author}{Chulliat, A.} \&
  \bibinfo{author}{Orgeval, J.-J.}
\newblock \bibinfo{title}{{Location of the North Magnetic Pole in April 2007}}.
\newblock \emph{\bibinfo{journal}{Earth, planets and space}}
  \textbf{\bibinfo{volume}{61}}, \bibinfo{pages}{703--710}
  (\bibinfo{year}{2009}).

\bibitem{Thebault_etal_2015a}
\bibinfo{author}{Th{\'e}bault, E.} \emph{et~al.}
\newblock \bibinfo{title}{International geomagnetic reference field: the
  twelfth generation}.
\newblock \emph{\bibinfo{journal}{Earth, Planets and Space}}
  \textbf{\bibinfo{volume}{67}} (\bibinfo{year}{2015}).

\bibitem{Friis-Christensen_etal_2006}
\bibinfo{author}{Friis-Christensen, E.}, \bibinfo{author}{L\"{u}hr, H.} \&
  \bibinfo{author}{Hulot, G.}
\newblock \bibinfo{title}{{Swarm: A constellation to study the Earth's magnetic
  field}}.
\newblock \emph{\bibinfo{journal}{Earth, planets and space}}
  \textbf{\bibinfo{volume}{58}}, \bibinfo{pages}{351--358}
  (\bibinfo{year}{2006}).

\bibitem{Gillet_etal_2015a}
\bibinfo{author}{Gillet, N.}, \bibinfo{author}{Barrois, O.} \&
  \bibinfo{author}{Finlay, C.~C.}
\newblock \bibinfo{title}{{Stochastic forecasting of the geomagnetic field from
  the COV-OBS.x1 geomagnetic field model, and candidate models for IGRF-12}}.
\newblock \emph{\bibinfo{journal}{Earth Planets Space}}
  \textbf{\bibinfo{volume}{67}}, \bibinfo{pages}{1321--14}
  (\bibinfo{year}{2015}).

\bibitem{Finlay_etal_2016}
\bibinfo{author}{Finlay, C.~C.}, \bibinfo{author}{Olsen, N.},
  \bibinfo{author}{Kotsiaros, S.}, \bibinfo{author}{Gillet, N.} \&
  \bibinfo{author}{Toeffner-Clausen, L.}
\newblock \bibinfo{title}{{Recent geomagnetic secular variation from {\it
  Swarm} and ground observatories as estimated in the CHAOS-6 geomagnetic field
  model}}.
\newblock \emph{\bibinfo{journal}{Earth Planets Space}}
  \textbf{\bibinfo{volume}{68}}, \bibinfo{pages}{1--18} (\bibinfo{year}{2016}).

\bibitem{Chulliat_etal_2019}
\bibinfo{author}{Chulliat, A.} \emph{et~al.}
\newblock \bibinfo{title}{{Out-of-Cycle Update of the US/UK World Magnetic
  Model for 2015-2020}}.
\newblock \bibinfo{type}{Tech. Rep.} (\bibinfo{year}{2019}).

\bibitem{Hope_1957}
\bibinfo{author}{Hope, E.~R.}
\newblock \bibinfo{title}{Linear secular oscillation of the northern magnetic
  pole}.
\newblock \emph{\bibinfo{journal}{Journal of Geophysical Research}}
  \textbf{\bibinfo{volume}{62}}, \bibinfo{pages}{19--27}
  (\bibinfo{year}{1957}).

\bibitem{Olsen_Mandea_2007}
\bibinfo{author}{Olsen, N.} \& \bibinfo{author}{Mandea, M.}
\newblock \bibinfo{title}{{Will the magnetic North Pole move to Siberia?}}
\newblock \emph{\bibinfo{journal}{Eos, Transactions American Geophysical
  Union}} \textbf{\bibinfo{volume}{88}}, \bibinfo{pages}{293--293}
  (\bibinfo{year}{2007}).

\bibitem{Korte_Mandea_2008}
\bibinfo{author}{Korte, M.} \& \bibinfo{author}{Mandea, M.}
\newblock \bibinfo{title}{{Magnetic poles and dipole tilt variation over the
  past decades to millennia}}.
\newblock \emph{\bibinfo{journal}{Earth Planets Space}}
  \textbf{\bibinfo{volume}{60}}, \bibinfo{pages}{937--948}
  (\bibinfo{year}{2008}).

\bibitem{Mandea_Dormy_2003}
\bibinfo{author}{Mandea, M.} \& \bibinfo{author}{Dormy, E.}
\newblock \bibinfo{title}{{Asymmetric behavior of magnetic dip poles}}.
\newblock \emph{\bibinfo{journal}{Earth Planets Space}}
  \textbf{\bibinfo{volume}{55}}, \bibinfo{pages}{153--157}
  (\bibinfo{year}{2003}).

\bibitem{Hansteen_1819}
\bibinfo{author}{Hansteen, C.}
\newblock \emph{\bibinfo{title}{{Untersuchungen \"{u}ber den magnetismus der
  erde}}} (\bibinfo{publisher}{Christiania, Gedruckt bey J. Lehmann und C.
  Gr\"{o}ndahl}, \bibinfo{year}{1819}).

\bibitem{Bloxham_Gubbins_85}
\bibinfo{author}{Bloxham, J.} \& \bibinfo{author}{Gubbins, D.}
\newblock \bibinfo{title}{The secular variation of {E}arth's magnetic field}.
\newblock \emph{\bibinfo{journal}{Nature}} \textbf{\bibinfo{volume}{317}},
  \bibinfo{pages}{777--781} (\bibinfo{year}{1985}).

\bibitem{Chulliat_etal_2010}
\bibinfo{author}{Chulliat, A.}, \bibinfo{author}{Hulot, G.} \&
  \bibinfo{author}{Newitt, L.~R.}
\newblock \bibinfo{title}{{Magnetic flux expulsion from the core as a possible
  cause of the unusually large acceleration of the north magnetic pole during
  the 1990s}}.
\newblock \emph{\bibinfo{journal}{J. Geophys. Res.}}
  \textbf{\bibinfo{volume}{115}}, \bibinfo{pages}{B07101}
  (\bibinfo{year}{2010}).

\bibitem{Gubbins_Roberts_83}
\bibinfo{author}{Gubbins, D.} \& \bibinfo{author}{Roberts, N.}
\newblock \bibinfo{title}{{Use of the frozen flux approximation in the
  interpretation of archaeomagnetic and palaeomagnetic data}}.
\newblock \emph{\bibinfo{journal}{Geophys J. Int.}}
  \textbf{\bibinfo{volume}{73}}, \bibinfo{pages}{675--687}
  (\bibinfo{year}{1983}).

\bibitem{Roberts_Scott_65}
\bibinfo{author}{Roberts, P.~H.} \& \bibinfo{author}{Scott, S.}
\newblock \bibinfo{title}{On analysis of secular variation. 1. {A} hydromagneic
  constraint - theory}.
\newblock \emph{\bibinfo{journal}{J Geomagn Geoelectr}}
  \textbf{\bibinfo{volume}{17}}, \bibinfo{pages}{137--151}
  (\bibinfo{year}{1965}).

\bibitem{Barrois_etal_2017}
\bibinfo{author}{Barrois, O.}, \bibinfo{author}{Gillet, N.} \&
  \bibinfo{author}{Aubert, J.}
\newblock \bibinfo{title}{Contributions to the geomagnetic secular variation
  from a reanalysis of core surface dynamics}.
\newblock \emph{\bibinfo{journal}{Geophysical Journal International}}
  \textbf{\bibinfo{volume}{211}}, \bibinfo{pages}{50--68}
  (\bibinfo{year}{2017}).

\bibitem{Barrois_etal_2018}
\bibinfo{author}{Barrois, O.}, \bibinfo{author}{Hammer, M.~D.},
  \bibinfo{author}{Finlay, C.~C.}, \bibinfo{author}{Martin, Y.} \&
  \bibinfo{author}{Gillet, N.}
\newblock \bibinfo{title}{{Assimilation of ground and satellite magnetic
  measurements: inference of core surface magnetic and velocity field
  changes}}.
\newblock \emph{\bibinfo{journal}{Geophys J. Int.}}
  \textbf{\bibinfo{volume}{215}}, \bibinfo{pages}{695--712}
  (\bibinfo{year}{2018}).

\bibitem{Barrois_etal_2019}
\bibinfo{author}{Barrois, O.} \emph{et~al.}
\newblock \bibinfo{title}{{Erratum: {\textquoteleft}Contributions to the
  geomagnetic secular variation from a reanalysis of core surface
  dynamics{\textquoteright} and {\textquoteleft}Assimilation of ground and
  satellite magnetic measurements: inference of core surface magnetic and
  velocity field changes{\textquoteright}}}.
\newblock \emph{\bibinfo{journal}{Geophys J. Int.}}
  \textbf{\bibinfo{volume}{216}}, \bibinfo{pages}{2106--2113}
  (\bibinfo{year}{2018}).

\bibitem{Livermore_etal_2017}
\bibinfo{author}{Livermore, P.~W.}, \bibinfo{author}{Hollerbach, R.} \&
  \bibinfo{author}{Finlay, C.~C.}
\newblock \bibinfo{title}{An accelerating high-latitude jet in {E}arth's core}.
\newblock \emph{\bibinfo{journal}{{Nature Geoscience}}}
  \textbf{\bibinfo{volume}{10}}, \bibinfo{pages}{62--68}
  (\bibinfo{year}{2017}).

\bibitem{Pais_Jault_2008}
\bibinfo{author}{Pais, A.} \& \bibinfo{author}{Jault, D.}
\newblock \bibinfo{title}{Quasi-geostrophic flows responsible for the secular
  variation of the {E}arth's magnetic field}.
\newblock \emph{\bibinfo{journal}{Geophys. J. Int.}}
  \textbf{\bibinfo{volume}{173}}, \bibinfo{pages}{421--443}
  (\bibinfo{year}{2008}).

\bibitem{Gillet_etal_2015b}
\bibinfo{author}{Gillet, N.}, \bibinfo{author}{Jault, D.} \&
  \bibinfo{author}{Finlay, C.~C.}
\newblock \bibinfo{title}{{Planetary gyre, time-dependent eddies, torsional
  waves, and equatorial jets at the Earth's core surface}}.
\newblock \emph{\bibinfo{journal}{J. Geophys. Res.}}
  \textbf{\bibinfo{volume}{120}}, \bibinfo{pages}{3991--4013}
  (\bibinfo{year}{2015}).

\bibitem{Aubert_2015}
\bibinfo{author}{Aubert, J.}
\newblock \bibinfo{title}{Geomagnetic forecasts driven by thermal wind dynamics
  in the earth's core}.
\newblock \emph{\bibinfo{journal}{Geophys J. Int.}}
  \textbf{\bibinfo{volume}{203}}, \bibinfo{pages}{1738--1751}
  (\bibinfo{year}{2015}).

\bibitem{Tangborn_Kuang_2018}
\bibinfo{author}{Tangborn, A.} \& \bibinfo{author}{Kuang, W.}
\newblock \bibinfo{title}{Impact of archeomagnetic field model data on modern
  era geomagnetic forecasts}.
\newblock \emph{\bibinfo{journal}{Physics of the Earth and Planetary
  Interiors}} \textbf{\bibinfo{volume}{276}}, \bibinfo{pages}{2 -- 9}
  (\bibinfo{year}{2018}).
\newblock \bibinfo{note}{Special Issue:15th SEDI conference}.

\bibitem{Sanchez_etal_2019}
\bibinfo{author}{Sanchez, S.}, \bibinfo{author}{Wicht, J.},
  \bibinfo{author}{B{\"a}renzung, J.} \& \bibinfo{author}{Holschneider, M.}
\newblock \bibinfo{title}{{Sequential assimilation of geomagnetic observations:
  perspectives for the reconstruction and prediction of core dynamics}}.
\newblock \emph{\bibinfo{journal}{Geophysical Journal International}}
  \textbf{\bibinfo{volume}{217}}, \bibinfo{pages}{1434--1450}
  (\bibinfo{year}{2019}).

\bibitem{Nilsson_etal_2014}
\bibinfo{author}{Nilsson, A.}, \bibinfo{author}{Holme, R.},
  \bibinfo{author}{Korte, M.}, \bibinfo{author}{Suttie, N.} \&
  \bibinfo{author}{Hill, M.}
\newblock \bibinfo{title}{{Reconstructing Holocene geomagnetic field variation:
  new methods, models and implications}}.
\newblock \emph{\bibinfo{journal}{Geophys J Int}}
  \textbf{\bibinfo{volume}{198}}, \bibinfo{pages}{229--248}
  (\bibinfo{year}{2014}).

\bibitem{Panovska_etal_2018}
\bibinfo{author}{Panovska, S.}, \bibinfo{author}{Constable, C.} \&
  \bibinfo{author}{Korte, M.}
\newblock \bibinfo{title}{Extending global continuous geomagnetic field
  reconstructions on timescales beyond human civilization}.
\newblock \emph{\bibinfo{journal}{Geochemistry, Geophysics, Geosystems}}
  \textbf{\bibinfo{volume}{19}}, \bibinfo{pages}{4757--4772}
  (\bibinfo{year}{2018}).

\bibitem{Metman_etal_2019}
\bibinfo{author}{Metman, M.~C.}, \bibinfo{author}{Livermore, P.~W.},
  \bibinfo{author}{Mound, J.~E.} \& \bibinfo{author}{Beggan, C.~D.}
\newblock \bibinfo{title}{Modelling decadal secular variation with only
  magnetic diffusion}.
\newblock \emph{\bibinfo{journal}{Geophysical Journal International}}
  \textbf{\bibinfo{volume}{219}}, \bibinfo{pages}{S58--S82}
  (\bibinfo{year}{2019}).

\end{thebibliography}

\clearpage

\section*{Acknowledgements}
PWL acknowledges funding from the Natural Environment Research Council (NERC) grant NE/P016758/1.
CCF acknowledges funding the the European Research Council (ERC) under the European Union's Horizon 2020 research and innovation programme, grant agreement No. 772561. 

\section*{Data availability}
The CHAOS-6-x8 and COV-OBS.x1 geomagnetic field models on which this study is based can be found at:\\ \verb!http://www.spacecenter.dk/files/magnetic-models/!\\
The flow models of Barrois et al. employed here can be found at:\\
\verb!https://geodyn.univ-grenoble-alpes.fr/!

\section*{Code availability}
All codes are freely available by request from P.W. Livermore~(email: p.w.livermore@leeds.ac.uk). 

\section*{Author contributions}
PWL and CCF devised the study; calculations were performed by PWL and MB. 
CCF derived the CHAOS-6-x8 field model. 
PWL and CCF analysed the geomagnetic field and core flow models, interpreted the results and wrote the paper.  All authors commented on the manuscript.

\section*{Author information}
\begin{itemize}
 \item The authors declare that they have no
competing financial interests.
\item Correspondence and requests for materials
should be addressed to P.W. Livermore~(email: p.w.livermore@leeds.ac.uk).
\end{itemize}

\begin{figure}
\begin{minipage}{8cm}
\includegraphics[height=20cm]{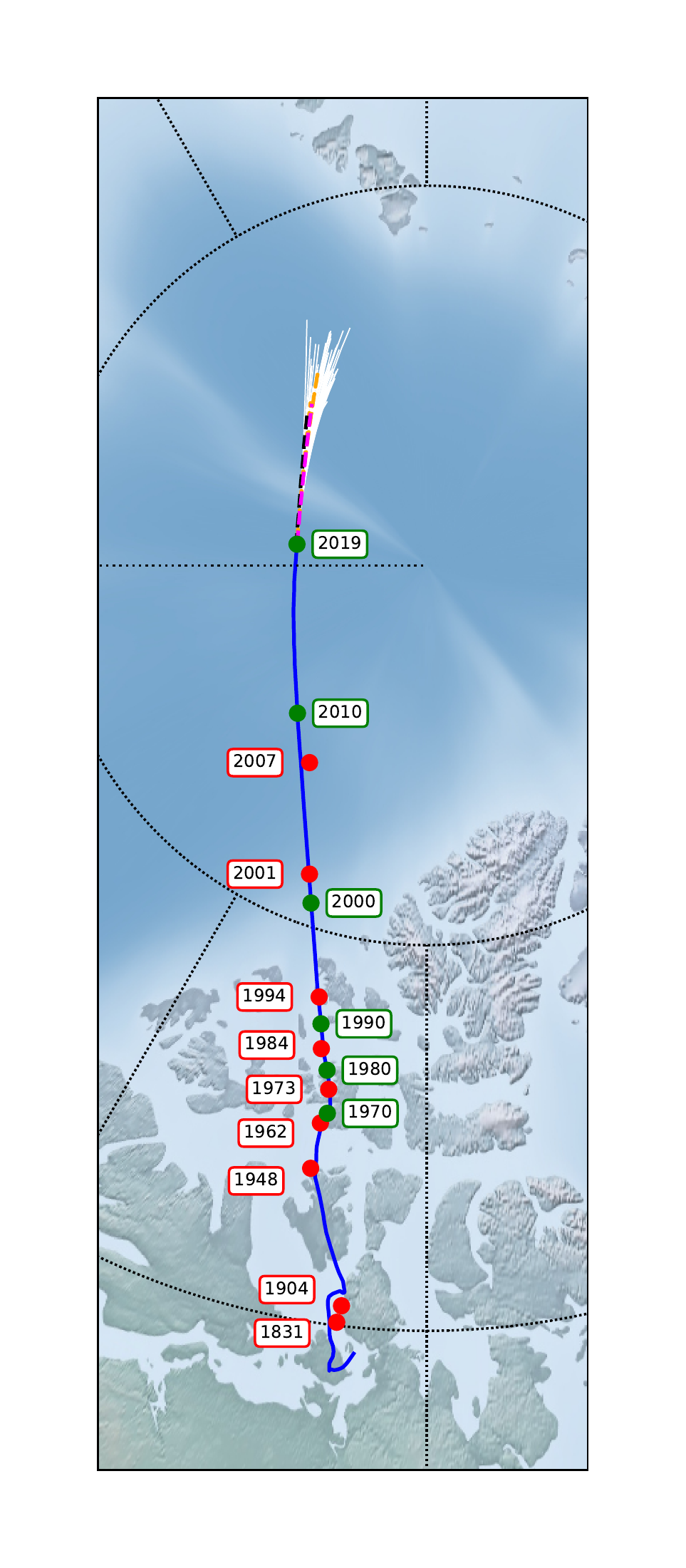}
\end{minipage}
\hfill
\begin{minipage}{8cm}

\caption{Historical movement and predicted future path of the North Magnetic pole in stereographic projection. Solid blue shows the pole's evolution according to the COV-OBS.x1 (1840-1998) and CHAOS-6-x8 (1999-2019) geomagnetic field models, with green circles indicating recent decadal positions;  red circles mark in-situ measurements (1831-2007) \cite{Mandea_Dormy_2003, Newitt_etal_2009}. The international date line is shown by the dotted black line on the 180$^\circ$ meridian. Predictions (see methods) 2019-2029 are: linear extrapolation from the World Magnetic Model v2 \cite{Chulliat_etal_2019} as black, linear extrapolation from CHAOS-6-x8 as magenta, a purely-diffusive model based on fitting geomagnetic secular variation over 2014-2019 
in orange \cite{Metman_etal_2019} and frozen-flux evolution 
using an ensemble of large-scale flows \cite{Barrois_etal_2018,Barrois_etal_2019} as white.}
\end{minipage}
\label{fig:summary}
\end{figure}

\begin{figure}
\includegraphics[width=15cm]{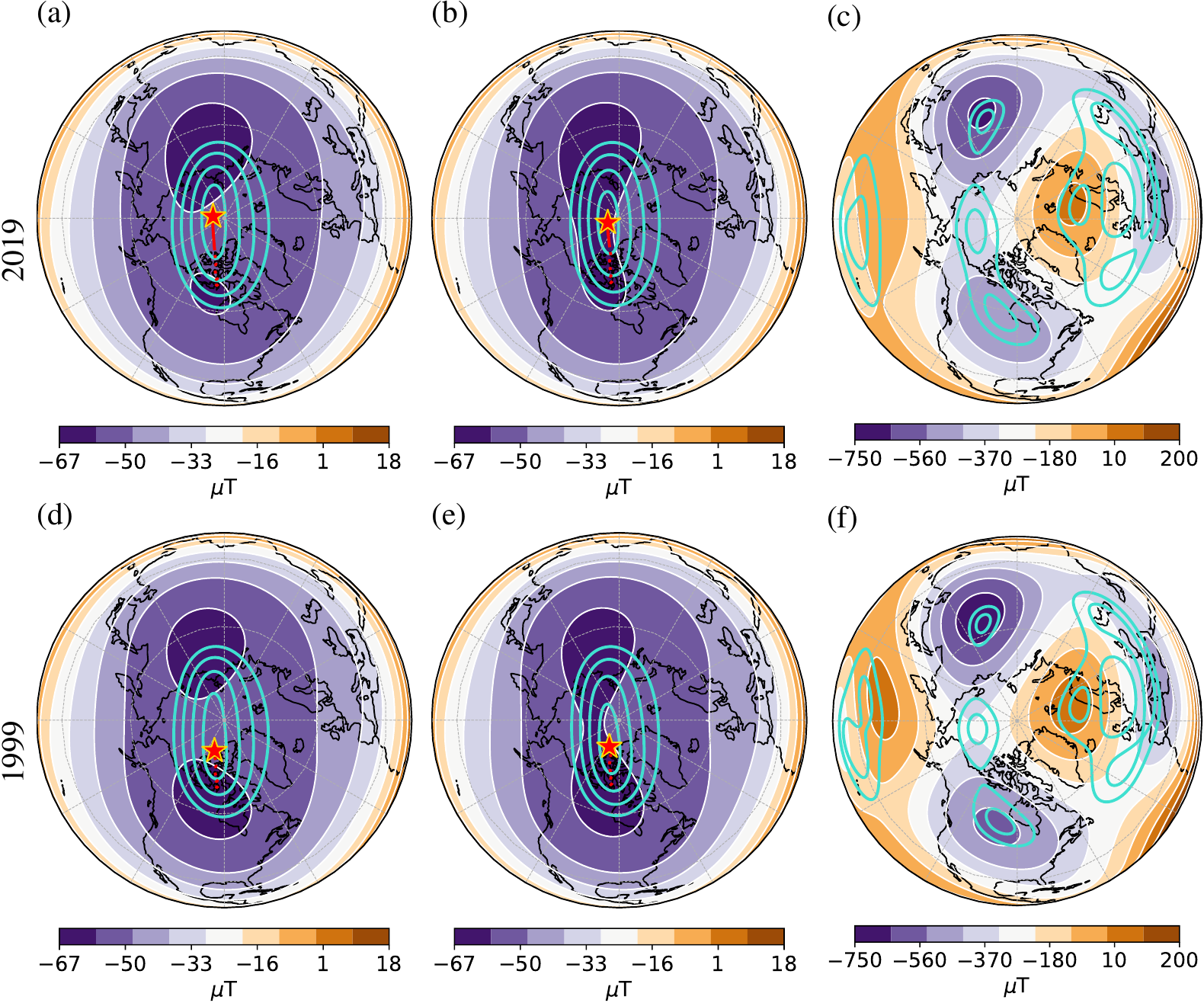}
\caption{A comparison of the structure of the geomagnetic field and the north magnetic pole position in orthographic projection between 2019 (a-c) and 1999 (d-f). 
(a,d): contours of the radial field on the Earth's surface overlaid with contours of H in turquoise (values [2,4,6,8] $\mu$T) and the north magnetic pole as a red star with its dotted tail showing the path 1840-1999, solid tail 1999-2019.
(b,e): as (a,d) but truncated to spherical harmonic degree 6. 
(c,f): structure of the geomagnetic field to degree 6 on the core-mantle-boundary, shown by contours of radial field overlaid with contours of H in turquoise (values [50,100] $\mu$T).}

\end{figure}

\begin{figure}

\includegraphics[width=15cm]{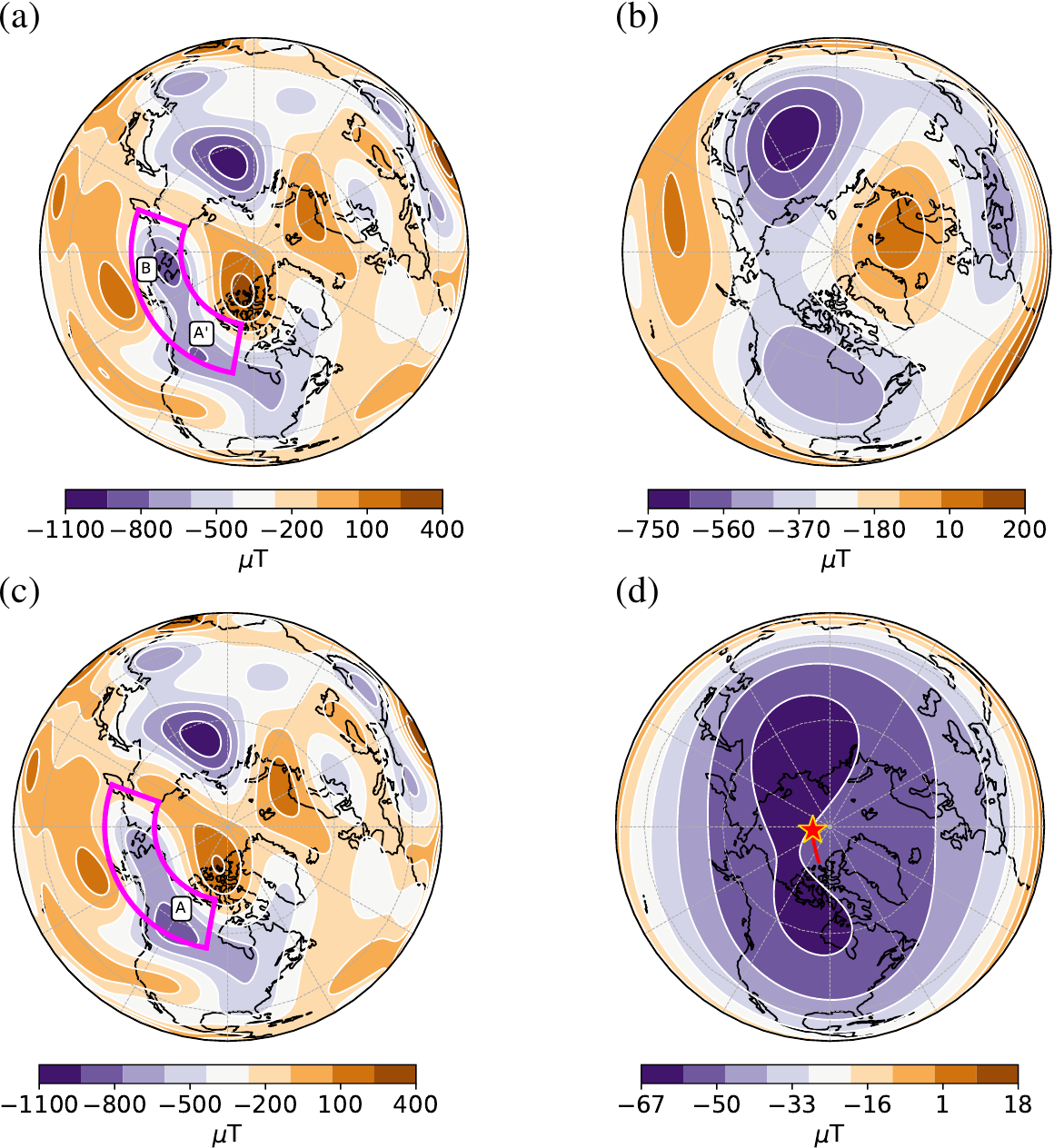}
\label{fig:fig3}
\caption{Experiment demonstrating the effect of elongation of the Canadian CMB flux lobe on the large-scale surface field and pole position. 
(c) contours of the radial component in 1999 according to CHAOS-6-x8.
(a) radial component of a composite field projected into a divergence-free spherical-harmonic representation, comprising the structure in 2019 within the magenta wedge and the structure in 1999 elsewhere; 
(b) radial field on the CMB, as in (a) but truncated to degree $6$, note the similar structure to Fig 2(c) demonstrates that flux lobe elongation explains the change in the Canadian surface patch;
(d) radial field on the Earth's surface with the north magnetic pole (red star), whose tail indicates its path since 1999, produced only by changes within the wedge.} 
\end{figure}

\begin{figure}
\includegraphics[width=15cm]{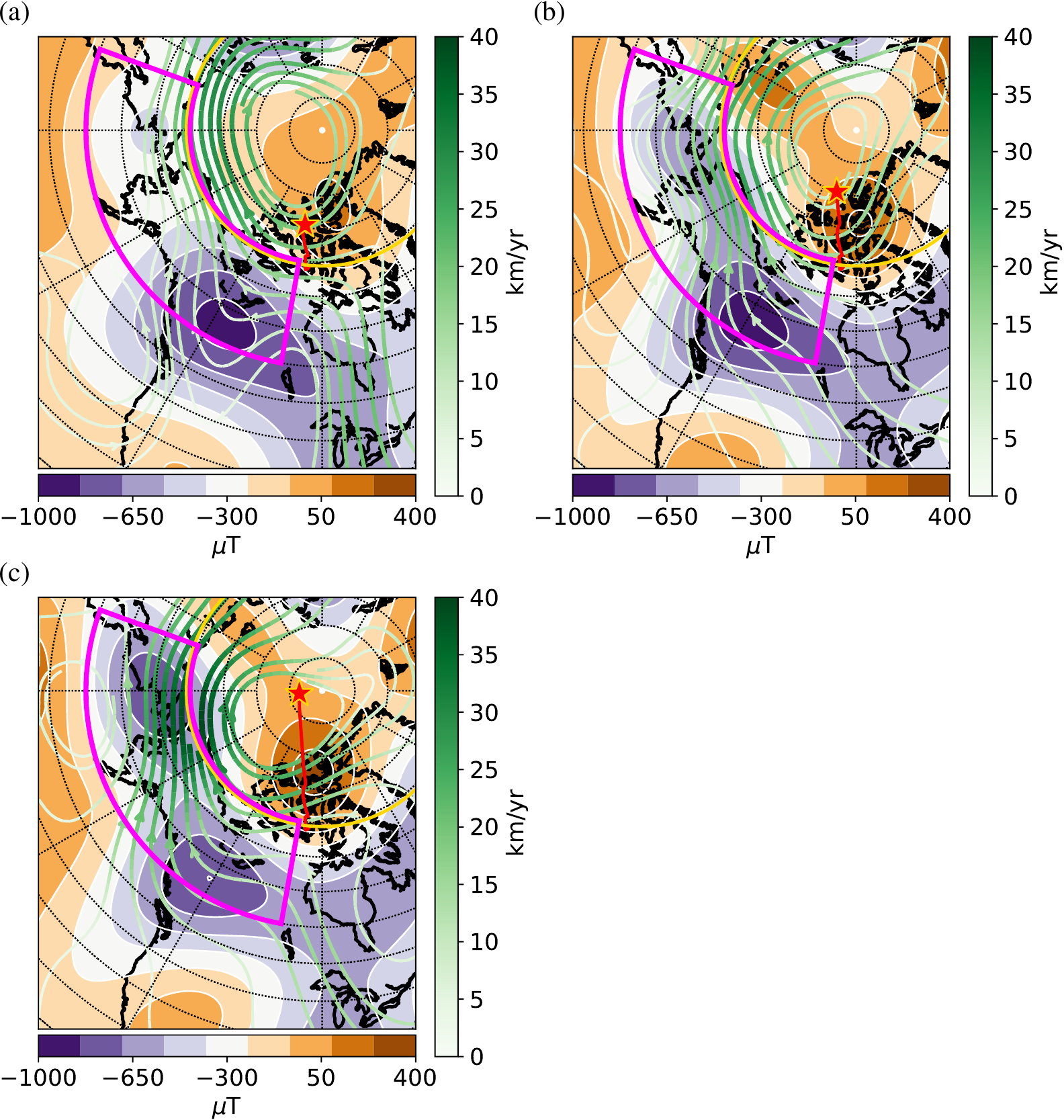}
\label{fig:fig4}
\caption{Local core surface dynamics around the Canadian flux lobe in stereographic projection at (a) 1970, (b) 1999 and (c) 2017, showing contours of the radial magnetic field, the north magnetic pole position and path since 1840, flow streamlines with arrows and the wedge within which flux lobe elongation occurs.
The 1970 magnetic field and flow data is from COV-OBS.x1 and the  ensemble mean flow of \cite{Barrois_etal_2017,Barrois_etal_2019}; those from 1999 are from CHAOS-6-x8 and the ensemble mean flow of \cite{Barrois_etal_2017,Barrois_etal_2019}; those from 2017 are from CHAOS-6-x8 and the ensemble mean of \cite{Barrois_etal_2018,Barrois_etal_2019}. The inner-tangent cylinder is marked in gold at about 69$^\circ$ N.
}
\end{figure}

\clearpage

\end{document}